\documentclass[aps,prl,twocolumn,showpacs,amsmath,amssymb]{revtex4}

\usepackage{times}

\usepackage{graphicx}
\usepackage{graphics}
\usepackage{longtable}
\usepackage{color}
\usepackage{soul}

\begin{document}

\noindent {\bf Published in Nature Materials (2012). \\
DOI: 10.1038/nmat3416. \\
URL: http://www.nature.com/nmat/journal/vaop/ncurrent/full/nmat3416.html} \\

\title{Hidden polymorphs drive the vitrification in ${\rm B_2O_3}$} 

\author{Guillaume Ferlat$^1$, Ari Paavo Seitsonen$^{1,2}$, Michele Lazzeri$^1$ \& Francesco Mauri$^1$}

\affiliation{$^1$IMPMC, CNRS$-$Universit\'e Pierre et Marie Curie, 4, Place Jussieu, F-75005 Paris, France},

\affiliation{$^2$ Physikalisch-Chemisches Institut der Universit\"at Z\"urich,
 Winterthurerstrasse 190, CH-8057, Switzerland.}

\date{\today}

\begin{abstract}

 Understanding the conditions which favor crystallisation or vitrification of liquids has been a long-standing scientific problem~\cite{cohe_61_1,bhat_2007_1,shin_2006_1}. Another connected, and not yet well understood question is the relationship between the glassy and the various possible crystalline forms a system may adopt~\cite{shen_2006_1,mart_2002_1}. 
In this context, ${\rm B_2O_3}$ is a puzzling case of study since i) it is one of the best glass-forming systems despite an apparent lack of low-pressure polymorphism ii) it vitrifies in a glassy form abnormally different from the only known crystalline phase at ambient pressure~\cite{youn_95_1} iii) it never crystallises from the melt unless pressure is applied, an intriguing behaviour known as the {\em crystallisation anomaly}~\cite{krac_38_1,uhlm_67_1,aziz_85_1}.
Here, by means of {\it ab-initio} calculations, we discover the existence of novel ${\rm B_2O_3}$ crystalline polymorphs with structural properties similar to the glass and formation energies comparable to the known ambient crystal. The resulting configurational degeneracy drives the system vitrification at ambient pressure. The degeneracy is lifted under pressure, unveiling the origin of the {\em crystallisation anomaly}. 
This work reconciles the behaviour of ${\rm B_2O_3}$ with that from other glassy systems and reaffirms the role played by polymorphism in a system's ability to vitrify~\cite{wang_76_1,good_75_1}.
Some of the predicted crystals are cage-like materials entirely made of three-fold rings, opening new perspectives for the synthesis of boron-based nanoporous materials.

\end{abstract}

\maketitle

 Polymorphism, the possibility for a substance to form several distinct crystalline phases of identical composition, is observed for a wide range of materials. This phenomenon has tremendous importance not only {\em per se} for understanding the crystallisation process but also because of practical implications such as the design and control of new materials with specific properties~\cite{desi_97_1}, a major issue for the pharmaceutical industry~\cite{llin_2008_1}. 
Indeed, the ability of molecular units to pack in various ways generates crystal phases which generally differ in their physical properties and cohesive energies.  
 Another important implication of polymorphism is related to the glassy state: as pointed out earlier~\cite{good_75_1,wang_76_1}, glass formation is often prevalent for those materials which are found in a variety of crystalline forms. An obvious example is silica (${\rm SiO_2}$), the archetypal glass-former, which at low pressure is found as quartz, cristobalite, keatite, tridymite, coesite and moganite~\cite{elli_90_1}. The existence of these many polytypes illustrate that the structural units, here the ${\rm SiO_4}$ tetrahedra, can occur in several conformations with little difference in strain energy, allowing for the possibility of metastable states~\cite{elli_90_1}. 
The ease of glass formation is usually understood as the result of the system frustration associated to the presence of many minima of comparable energy in the crystal energy landscape (CEL). 
In the context of organic chemistry, it has also been observed that systems with many almost equi-energetic structures containing a common interchangeable motif, correlate with a tendency to disorder~\cite{pric_2009_1}. This is the case for instance for Chlorothalonil or Aspirin. On the contrary, 2,3-dichloronitrobenzene shows a monomorphic behavior because of the large energy difference ($>$ 7 kcal.mol$^{-1}$) between the most stable form and its other polytypes~\cite{pric_2009_1}. 

The existence of a correlation between a system's ability to amorphise and the presence of numerous crystal polymorphs with comparable energies is remarkable since it seemingly applies to a very broad range of materials, which includes organic, inorganic and metallic systems. 
In addition, several studies have pointed out a structural link between the glassy and the known crystalline phases~\cite{mart_2002_1,shen_2006_1}. These connections may also have ramifications for the understanding of phenomena such as polyamorphism (existence of distinct amorphous states), liquid-liquid transitions and even possibly for protein folding~\cite{sali_94_1}.
  
Boron oxide (${\rm B_2O_3}$) is one of the simplest glass-forming oxides and has received considerable attention~\cite{youn_95_1,lee_2005_2,umar_2005_1,ferl_2008_1,huan_2006_1,wrig_2010_1}. It is one of the best glass formers, even a better one than silica, since it has never been observed to crystallize from a dry melt at ambient pressure: even if the melt is seeded with crystals and maintained for several months at various temperatures below the melting point, no crystal growth is observed at any imposed cooling rates ($|dT/dt|>10^{-5}$ K.s$^{-1}$)~\cite{krac_38_1,uhlm_67_1,aziz_85_1}. Spontaneous crystallisation from the melt is only obtained when the pressure is raised above a threshold level, typically in the range 0.4-1.0 GPa~\cite{uhlm_67_1,aziz_85_1}. This puzzling behavior has been termed the ${\rm B_2O_3}$ {\it crystallisation anomaly}~\cite{uhlm_67_1,aziz_85_1}. Although nucleation theory based models have been put forward to account for this behavior~\cite{aziz_85_1}, its microscopic origin remains unexplained.

For pressure below 2.0 GPa, the obtained crystal, made of ribbons of ${\rm BO_3}$ planar triangle units, is ${\rm B_2O_3}$-I. At higher pressure, another polymorph, ${\rm B_2O_3}$-II, in which the boron atoms are four-fold coordinated is crystallised~\cite{uhlm_67_1}.
Alternatively, crystalline ${\rm B_2O_3}$-I can also be prepared by the stepwise dehydration of orthoboric acid (${\rm H_3BO_3}$)\cite{mccu_37_1} or by seeding a melt with borophosphate~\cite{klin_68_1}. 

The glassy state has also remarkable {\em anomalies}:
1) with only two experimentally known polytypes, well separated in energy, the ease of vitrification of ${\rm B_2O_3}$ seems at odds with the correlation mentioned before: a poor polymorphism should reflect in a poor glass-former;
2) the glass density (1.84 g.cm$^{-3}$) is significantly lower than that of the known crystalline polymorphs, by a magnitude of 30 and 40 \% (2.55 g.cm$^{-3}$ for ${\rm B_2O_3}$-I and 3.11 g.cm$^{-3}$ for ${\rm B_2O_3}$-II);
3) the glass medium-range order is markedly different from that of the crystalline polymorphs: in the glass, a majority of the ${\rm BO_3}$ molecular units form three-fold rings, ${\rm B_3O_6}$, known as boroxol rings (Fig. 1)~\cite{youn_95_1,umar_2005_1,ferl_2008_1}. Strangely, these rings, which can be viewed as superstructural units, are totally absent in both ${\rm B_2O_3}$ crystalline polymorphs. Boroxol rings are however found in large amounts in certain borate crystals such as ${\rm K_3B_3O_6}$ or ${\rm Cs_2O}$-9${\rm B_2O_3}$~\cite{wrig_2010_1}. 

\begin{figure}[t!]
\includegraphics[bb=20 20 730 550,scale=0.34]{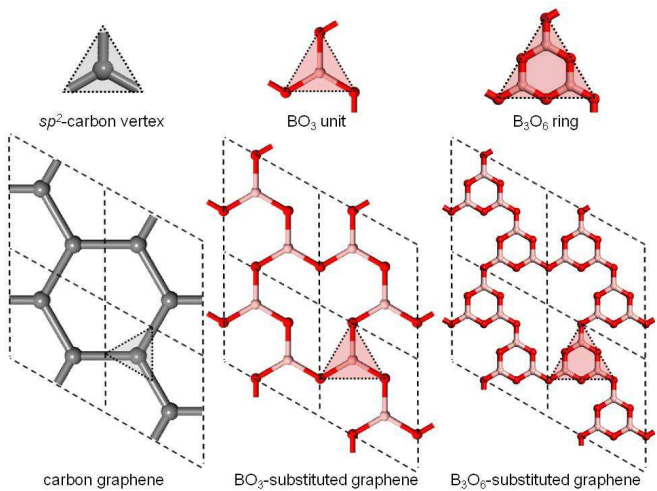}
\caption{\label{fig_1} Construction of new ${\rm B_2O_3}$ polymorphs: Vertices of $sp^2$-carbon structures are replaced either by the ${\rm BO_3}$ triangle or the ${\rm B_3O_6}$ (boroxol) ring as illustrated here in the case of a graphene layer. The T0 and T0-$b$ polymorphs were obtained by stacking the ${\rm BO_3}$- and ${\rm B_3O_6}$-substituted layers respectively.}
\vspace{-0.5cm}
\end{figure}

The seemingly poor polymorphism of ${\rm B_2O_3}$ compared to silica and the fact that pressure is required to crystallize ${\rm B_2O_3}$-I from the anhydrous melt raise the question whether ${\rm B_2O_3}$-I is indeed the equilibrium crystalline phase at ambient pressure. 
Although unknown polymorphs have been proposed in the literature~\cite{taka_2003_1,huan_2006_1}, none of them can explain consistently the mentioned anomalies since the predicted structures appear at too high energies in the CEL.
This motivated the present systematic exploration of the ${\rm B_2O_3}$ CEL.

This was done by exploiting the isomorphism between the trigonal network of ${\rm B_2O_3}$ and other three-fold coordinated networks such as $sp^2$-carbon structures (see Methods). 
We investigated all possible networks (13) with up to six vertices in the unit cell, as obtained from an exhaustive search originally applied to carbon polymorphs~\cite{wink_2001_1}. We also considered a layered structure based upon the topology of graphite (Fig. 1). By placing ${\rm BO_3}$ triangle units at the vertices of the original structures, 13 topologically different ${\rm B_2O_3}$ polymorphs were generated which are labeled T0 to T13 in the following. Among them, T13 is the known ${\rm B_2O_3}$-I crystal. In T8 and T10, 50 \% of the boron atoms belong to three-fold rings, i.e. boroxol rings.  
Further to expand the search and to investigate the role of the boroxol ring as a structural motif, 13 additional structures, labeled T0-$b$ to ${\rm B_2O_3}$-I-$b$, were generated by replacing the ${\rm BO_3}$ units by the ${\rm B_3O_6}$ ones. In this way, structures which are made of 100 \% boroxol units were obtained, taking advantage of the self-similarity between a ${\rm BO_3}$ and a ${\rm B_3O_6}$ unit (Fig. 1).
All structures were then relaxed using {\em first-principles} calculations within the density functional theory framework.

The resulting ${\rm B_2O_3}$ polymorphs have low (1.0-2.0 g.cm$^{-3}$) to very low ($<$ 0.7 g.cm$^{-3}$) densities.
Most of the new crystals are microporous due to cage- or channel-like structures. The shape of these channels is either circular, rectangular or triangular with a minimum length of typically 6 to 17 \AA \ (see Fig. 2 for an example).

\begin{figure}
\includegraphics[bb=20 20 370 550,scale=0.45]{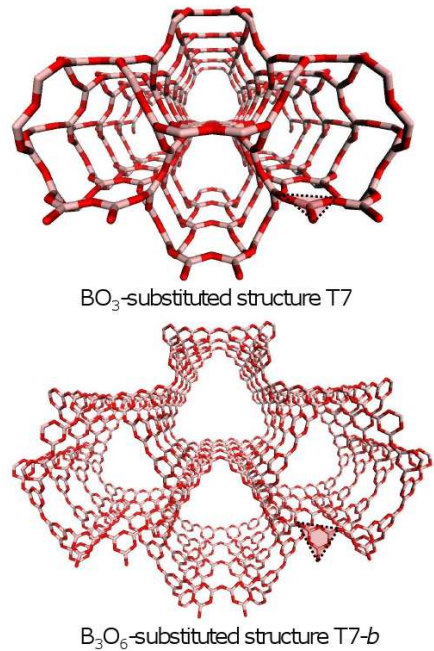}
\caption{\label{fig_2} Examples of nanoporous structures: The ${\rm BO_3}$-made T7 and its corresponding ${\rm B_3O_6}$-made T7-$b$ structures: large channels, of diameter approximately 8.0 \AA \ and 11.0 \AA \ respectively, are clearly seen.}
\vspace{-0.5cm}
\end{figure}

Fig. 3a shows the cohesive enthalpy versus density for all the obtained crystals. Very strikingly, many of them fall in a narrow energy range which includes the known ${\rm B_2O_3}$-I polymorph. This situation, with many minima in the CEL, is expected to prevent the nucleation of a given crystal by favoring a disordered structure instead. 
Therefore, this explains the {\em glass anomaly} 1) (ease of vitrification).

\begin{figure*}[]
\includegraphics[bb=20 20 800 620,scale=0.325]{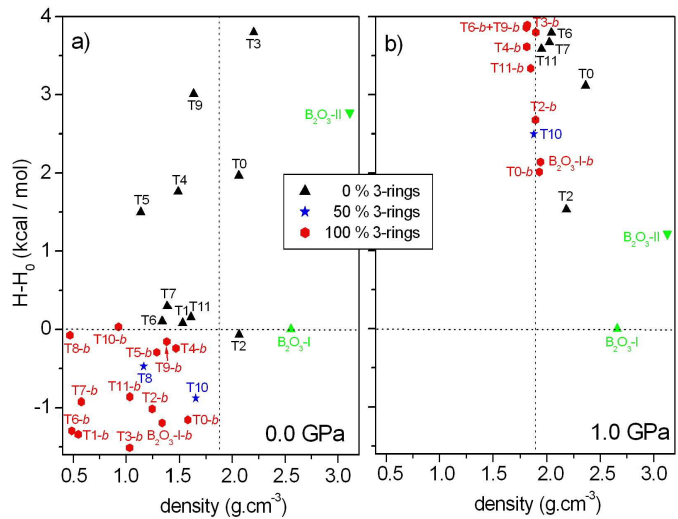}
\caption{\label{fig_3} Enthalpy as a function of the density for the obtained ${\rm B_2O_3}$ polymorphs: a) at P= 0.0 GPa b) at P = 1.0 GPa, on the same scale as a). The enthalpy reference, ${\rm H_0}$, is that obtained for ${\rm B_2O_3}$-I and is shown by the horizontal dashed line. The vertical dashed line indicates the glass density. The different symbols and colors refer to the proportion of boroxol rings in the polymorphs.
}
\vspace{-0.5cm}
\end{figure*}

All new predicted structures have densities lower than ${\rm B_2O_3}$-I and the majority of them fall in the range 1.1-1.7 g.cm$^{-3}$, i.e. close to the liquid density above and near the glass transition temperature ($\rho_{liquid}(2000\ K) \sim$ 1.5 g.cm$^{-3}$ and $\rho_{glass}(540\ K) \sim$ 1.8 g.cm$^{-3}$). 
This is consistent with the observed amorphisation in a low density glass as the temperature is decreased from the liquid state.
This thus explains the {\em glass anomaly} 2).

Another result of noteworthy relevance is the fact that the majority of the low energy polymorphs are those incorporating large amounts (50 or 100 \%) of boroxol rings. Indeed, the ${\rm BO_3}$ to ${\rm B_3O_6}$ substitution tend to give polymorphs with lower enthalpies (by $\sim $ 2.0 kcal/mol on average, up to $\sim$ 5.5 kcal/mol in the T3 case) with only two exceptions, T8 and T10, which already incorporate 50 \% of three-fold rings in the starting structures.
Interestingly, ${\rm B_2O_3}$-I-$b$, the boroxol-made equivalent of ${\rm B_2O_3}$-I, is found among the lowest energy polytypes. These findings, thus, confirm that boroxol rings are essential to stabilize low density structures~\cite{ferl_2008_1} and explain the {\em glass anomaly} 3).
We note that in the T10 structure, the dihedral angle distribution between adjacent ring and non-ring units is well in line with the one measured in the glass by two-dimensional NMR\cite{joo_2000_1}: this supports the idea that the glass shares structural similarities with the predicted polymorphs in the region of energy-density degeneracy. 

The enthalpy penalty with respect to pressure is larger for the new polymorphs than it is for ${\rm B_2O_3}$-I as a consequence of their low densities.  
As a result, the new polymorphs become less stable than ${\rm B_2O_3}$-I at relatively low pressures (Fig. 3b). The enthalpy separation between ${\rm B_2O_3}$-I and the second lowest polymorph rapidly increases with pressure ($\sim$ 0.8 kcal/mol at 0.5 GPa and $\sim$ 1.3 kcal/mol at 1.0 GPa), gradually leading to a situation in which a monomorphic behavior is expected~\cite{pric_2009_1}. 
This is fully in line with the experimental observation that a threshold pressure (0.4-1.0 GPa) is required to observe ${\rm B_2O_3}$-I crystallisation from the melt~\cite{krac_38_1,uhlm_67_1,aziz_85_1}.

Thus, the present results unravel the mystery of the {\em crystallisation anomaly}: in ${\rm B_2O_3}$, the crystallisation is avoided at ambient pressure as a result of the existence of several competing phases which eventually induces the system amorphisation. The degeneracy of ${\rm B_2O_3}$-I with the competing phases can be lifted by the application of a modest pressure which leaves ${\rm B_2O_3}$-I sufficiently separated in energy from other competitors. 

These results reveal a much richer polymorphism in ${\rm B_2O_3}$ than expected before. In this sense, the behavior of ${\rm B_2O_3}$ appears now to be much more similar to that from other well-known glassy systems such as silica.
The slight differences in energy between the polytypes comes from the variability of the B-O-B angle linking the structural units, just as in silica Si-O-Si angles bridging the tetrahedra.

An important issue raised by the present work is whether the predicted ${\rm B_2O_3}$ crystals can be experimentally synthesised and if so, how. All the structures investigated fall within the energy range of thermodynamically accessible polymorphs (a debatable quantity, of order of 3 kcal.mol$^{-1}$)~\cite{pric_2009_1}. Of course, kinetical or experimental difficulties may however prevent their observation. 
We note that there are at least two reports in the literature of low density polymorphs\cite{cole_35_1,koca_96_1}, both obtained by dehydration of ${\rm H_3BO_3}$. In the first one\cite{cole_35_1}, made in the early attempts to obtain ${\rm B_2O_3}$-I, the structure is a cubic phase of density 1.805 g.cm$^{-3}$. 
However, in this case, the fact that others have not been able to repeat the experiment remains puzzling. Part of the difficulty may well be due to the fact that in this density region, one needs to avoid the glass transition resulting from the polymorphic degeneracy in the energy-density diagram.
The second\cite{koca_96_1}, much more recent, reports an hexagonal phase of density 0.69 g.cm$^{-3}$, incorrectly assigned to ${\rm B_2O_3}$-I and which could well be one of our very low density polymorphs ($<$ 0.7 g.cm$^{-3}$, Fig. 3a). In this case, the low dehydration rate used\cite{koca_96_1} was most probably a key ingredient.   

In light of our predictions, further experimental studies dedicated to crystal synthesis would be very much indicated. Among the most promising directions towards the synthesis of low-density structures are sol-gel techniques~\cite{lang_96_1}. Starting from an alkyl (R) substituted metaboric acid, ${\rm R_3B_3O_6}$, boroxol-boroxol linkages could be formed via hydrolysis followed by either alcohol or water condensation~\cite{wrig_2010_1}.
 
Our results confirm that the presence of small rings is a very desirable feature to stabilize low density frameworks~\cite{ferl_2008_1}, a correlation which was first observed in the context of zeolites~\cite{brun_89_1}. 
The propensity of boron atoms to form three-membered rings is therefore quite appealing as illustrated by the recent discovery of crystalline covalent organic frameworks (COF), a novel family of high porosity and low density materials~\cite{cote_2005_1}: COF incorporate small rings such as boroxines, i.e. boroxol rings terminated by a radical such as alkyl, alkoxy or triaryl. 
Thus, our work opens interesting new perspectives for the search and design of boron-based nanoporous materials.

\paragraph*{Methods}
Starting from all possible $sp^2$-carbon 3D framework structures with up to six atoms in the primitive unit cell\cite{wink_2001_1}, we generated the T1-T14 structures by replacing the carbon by boron atoms and placing one oxygen in the middle of each B-B bond. The structure numbering corresponds to the order of appearance in table 1 of Ref. \onlinecite{wink_2001_1}. We rejected the structure T12, corresponding to the 6(3)5-09 net in Ref. \onlinecite{wink_2001_1}, since the carbon were almost four-coordinated. The structure T14 appeared to be identical to T6 and was not considered further. In addition, we constructed a layered structure, T0, based on the topology of graphite (Fig. 1). The structures T0-$b$ to T13-$b$ were obtained from T0-T13 by doubling each lattice constant and by substituting each ${\rm BO_3}$ unit for a ${\rm B_3O_6}$ unit. Depending on the polymorph, the primitive cell contains from 2 to 27 ${\rm B_2O_3}$ units (10 to 135 atoms). T13 is the known ${\rm B_2O_3}$-I polymorph, and thus, T13-$b$ is labeled ${\rm B_2O_3}$-I-$b$ in Fig. 3.
For each structure, both the atomic positions and the lattice cell were optimised using density functional theory calculations. Preliminary calculations, employing molecular dynamics simulations at 500 K were first carried out with the {\sc Siesta}\index{Siesta@{\sc Siesta}} code\cite{sole_2002_1}, norm-conserving pseudopotentials and localised basis sets as developed in Ref. \onlinecite{ferl_2006_2}.
Then, the obtained structures were relaxed at T = 0 K with the CASTEP plane-wave code~\cite{clar_2005_1}, ultrasoft pseudopotentials~\cite{vand_90_1} and the Perdew-Burke-Ernzerhof generalised gradient approximation functional~\cite{perd_96_1}. We used a plane-wave basis-set cutoff of 544 eV and a $k$-point grid spacing of 2$\pi\times$ 0.05 \AA$^{-1}$ or finer. Tolerances on the atomic forces and stress were set to 0.005 eV/\AA \ and 0.01 GPa, respectively. Because of the use of the PBE functional, the equilibrium densities are expected to be slightly underestimated. Therefore, the densities shown in Fig. 3 have been rescaled by a factor 1.109, as given by the ratio of the experimental (2.55 g.cm$^{-3}$) and calculated (2.30 g.cm$^{-3}$) densities of ${\rm B_2O_3}$-I.
For some of the networks investigated, several enthalpy minima corresponding to different symmetries were found. Only the lowest minimum of each topologically unrelated structure is shown in Fig. 3. For this reason, we do not show the enthalpy of ${\rm B_2O_3}$-0, a recently predicted crystal\cite{huan_2006_1} of different symmetry but the same topology as ${\rm B_2O_3}$-I. Other polymorphs previously proposed in the literature appear at higher energies~\cite{taka_2003_1} than the upper scale of Fig. 3.

\paragraph*{Acknowledgments}
This work was performed using HPC resources from GENCI-CINES/IDRIS (Grant x2010081875). We thank Ph. Depondt and E. Lacarce for critical reading of the manuscript. This paper is dedicated to the memory of J.-D. Pinou.

\end{document}